\documentclass[a4paper,11pt]{article}
\usepackage{pos}
\usepackage{epstopdf}
\usepackage{mathrsfs}

\title{Observation of ELVES with Mini-EUSO telescope on board the International Space Station}
 \ShortTitle{Mini-EUSO observation of ELVES}

\author*[a]{Laura Marcelli}
\author[b,c]{Enrico Arnone}
\author[b,c]{Matteo Barghini}
\author[b,c]{Matteo Battisti}
\author[d]{Alexander Belov}
\author[b,c]{Mario Bertaina}
\author[e]{Carl Blaksley}
\author[f]{Karl Bolmgren}
\author[g,a]{Giorgio Cambi\`e}
\author[h]{Francesca Capel}
\author[a,e,g]{Marco Casolino}
\author[e]{Toshikazu Ebisuzaki}
\author[f]{Christer Fuglesang} 
\author[e]{Philippe Gorodetzki} 
\author[i]{Fumiyoshi Kajino} 
\author[d]{Pavel Klimov}
\author[l]{Wlodzimierz Marsza{\l}}
\author[b,c]{Marco Mignone} 
\author[m]{Etienne Parizot} 
\author[g,a]{Piergiorgio Picozza} 
\author[n]{Lech Wictor Piotrowski} 
\author[b,c,l]{Zbigniew Plebaniak} 
\author[m]{Guilliame Pr\'ev\^ot} 
\author[g]{Giulia Romoli}
\author[g,a]{Enzo Reali}
\author[o]{Marco Ricci} 
\author[e]{Naoto Sakaki} 
\author[l]{Kenji Shinozaki} 
\author[l]{Jacek Szabelski} 
\author[e]{Yoshiyuki Takizawa}

\affiliation[a]{ INFN, Sezione di Roma Tor Vergata - Rome, Italy}
\affiliation[b]{ INFN, Sezione di Torino - Torino, Italy}
\affiliation[c]{ Dipartimento di Fisica, Universit\'a di Torino, Italy}
\affiliation[d]{ Skobeltsyn Institute of Nuclear Physics, Lomonosov Moscow State Univ. - Moscow, Russia}
\affiliation[e]{ RIKEN - Wako, Japan}
\affiliation[f]{ KTH Royal Institute of Technology - Stockholm, Sweden}
\affiliation[g]{ Universit\'a degli Studi di Roma Tor Vergata - Dipartimento di Fisica, Rome, Italy}
\affiliation[h]{ Technical University of Munich - Munich, Germany} 
\affiliation[i]{ Konan University, Kobe, Japan}
\affiliation[l]{ National Centre for Nuclear Research - Lodz, Poland}
\affiliation[m]{ Universit\'e de Paris, CNRS, Astroparticule et Cosmologie, F-75006 Paris, France}
\affiliation[n]{ Faculty of Physics, University of Warsaw - Warsaw, Poland}
\affiliation[o]{ INFN-LNF - Frascati, Italy}

\forColl{JEM--EUSO}

\emailAdd{marcelli@roma2.infn.it}

\abstract{Mini-EUSO is a detector observing the Earth in the ultraviolet band from the International Space Station through a nadir-facing window, transparent to the UV radiation, in the Russian Zvezda module. Mini-EUSO main detector consists in an optical system with two Fresnel lenses and a focal surface composed of an array of 36 Hamamatsu Multi-Anode Photo-Multiplier tubes, for a total of 2304 pixels, with single photon counting sensitivity. The telescope also contains two ancillary cameras, in the near infrared and visible ranges, to complement measurements in these bandwidths.
The instrument has a field of view of 44$^{\circ}$, a spatial resolution of about 6.3 km on the Earth surface and of about 4.7 km on the ionosphere.

The telescope detects UV emissions of cosmic, atmospheric and terrestrial origin on different time scales, from a few $\mu$s upwards. On the fastest timescale of 2.5 $\mu$s, Mini-EUSO is able to observe atmospheric phenomena as Transient Luminous Events and in particular the ELVES, which take place when an electromagnetic wave generated by intra-cloud lightning interacts with the ionosphere, ionizing it and producing apparently superluminal expanding rings of several 100 km and lasting $\simeq$ 100 $\mu$s. These highly energetic fast events have been observed to be produced in conjunction also with Terrestrial Gamma-Ray Flashes and therefore a detailed study of their characteristics (speed, radius, energy ...) is of crucial importance for the understanding of these phenomena.

In this paper we present the observational capabilities of ELVE detection by Mini-EUSO and specifically the reconstruction and study of ELVE characteristics.}

\FullConference{37$^{\rm{th}}$ International Cosmic Ray Conference (ICRC 2021)\\
		July 12th -- 23rd, 2021\\
		Online -- Berlin, Germany}

\begin{document} 
\maketitle

\section{Introduction}
\begin{figure}[ht]
\centering
\includegraphics[width=0.47\textwidth]{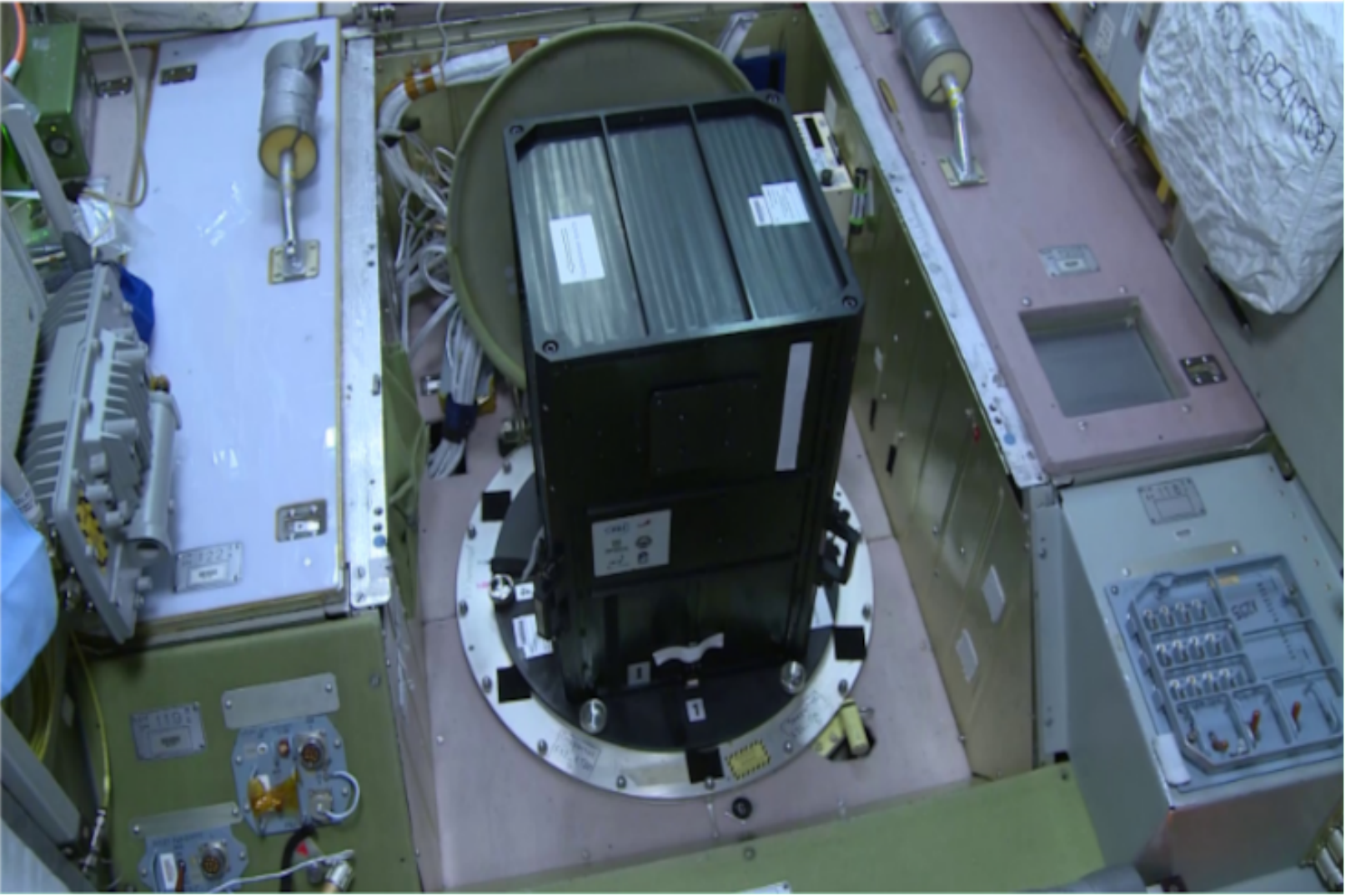}
\caption{Mini-EUSO installed inside the ISS on the UV transparent window of the Zvezda module.}
\label{instrument}        
\end{figure}

Mini-EUSO (Multiwavelength Imaging New Instrument for the Extreme Universe Space Observatory) \cite{Mini-EUSO-Astrophys}, also known as  'UV atmosphere' in the Russian Space Program, is a 2304 pixel telescope operating in the UV range (290 $\div$ 430 nm) and with a  field of view of $\simeq $44$^{\circ}$. It  was installed on the International Space Station (ISS) in 2019, with observations taking place from the Earth (nadir)-facing UV transparent window in the Zvezda module (Figure \ref{instrument}).   

The optical system consists of two Fresnel lenses with a diameter of 25 cm. The focal surface, or Photon Detector Module (PDM), is composed of 36 MultiAnode Photomultipliers (MAPMTs) tubes by Hamamatsu, 64 pixels each, capable of single photon detection. Readout is handled by a SPACIROC-3 ASICs with a sampling speed of  $2.5\: \mu s$ (1 Gate Time Unit, GTU). Data are then processed by a Zynq based FPGA board \cite{capel} which implements a multi-level triggering \cite{matteo}, allowing the measurement of triggered UV transients for 128 frames at time scales of both $2.5\: \mu$s and $320\: \mu$s. A continuous acquisition mode with 40.96 ms frames is also performed.  
In addition to the main detector, Mini-EUSO contains two ancillary cameras in visible ($400 \div 780$ nm) and near infrared ($1500 \div 1600$ nm) ranges to complement the UV  measurements and some single pixel UV sensors used to manage the day/night transition \cite{giorgio}. The telescope has a power consumption of 60 W, a weight of 35 kg and its dimensions are $37 \times 37 \times 62 \ cm^3$. The detector was integrated in the uncrewed Soyuz capsule and launched on 2019/08/22, with periodic acquisition sessions taking place since October 2019 \cite{casolino}.

The main scientific objectives of the mission are the search for nuclearites and Strange Quark Matter \cite{lech}, the study of atmospheric phenomena such as Transient Luminous Events, meteors and meteoroids, the observation of natural (such as clouds or marine bioluminescence) and artificial (such as city lights) nighttime terrestrial UV emissions \cite{kenji, alessio} and of artificial satellites and man-made space debris. It is also capable  of observing Extensive Air Showers generated by Ultra-High Energy Cosmic Rays with an energy above 10$^{21}$ eV and detecting artificial showers generated with lasers from the ground \cite{francesco}. 


\section{Observations of ELVES}\label{Sci_goals}
 For its high sampling speed (2.5 $\mu$s) and spatial resolution ($\simeq 4.7$  km at the ionosphere altitude (90 km)), Mini-EUSO is well suited to observe Transient Luminous Events (TLEs), upper atmospheric optical phenomena of electromagnetic nature, connected with thunderstorms. In this work we discuss the observation of ELVEs (Emission of Light and Very low frequency perturbations due to Electromagnetic pulse Sources). ELVES  were  predicted \citep{inan91}  before observation \citep{doi:10.1029/91GL03168, Fukunishi1996}, and last about a  few hundred $\mu s$.

\begin{figure}[ht]
\centering
\includegraphics[width=0.55\textwidth,page=40]{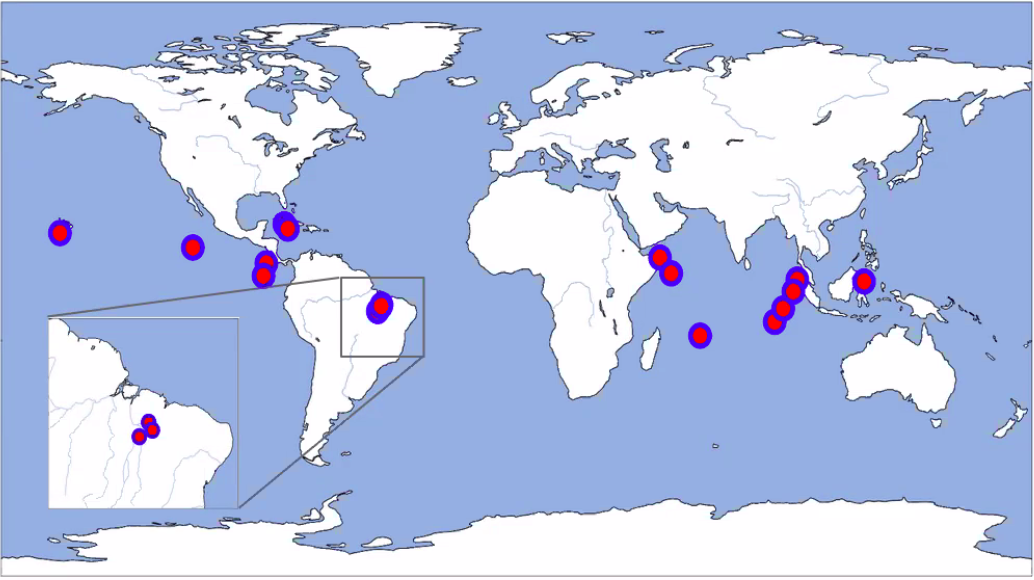}
\caption{Map of the 17 ELVEs detected with Mini-EUSO. ELVEs are distibuited mainly along the equatorial region.} 
\label{fig-ELVE-map}        
\end{figure}

Up to now 17 ELVES have been observed with Mini-EUSO in the first year of data, mostly in the low latitude regions (Figure \ref{fig-ELVE-map}), including three double-ringed ELVEs and one three-ringed ELVE.  

Mini-EUSO observes ELVES in the 2.5 $\mu$s time scale as large ring-like upper atmospheric emissions that appear to be expanding at superluminal speed and last $\simeq$ 100 $\mu$s. 

\begin{figure}[ht]
\centering
\includegraphics[width=1.0\textwidth]{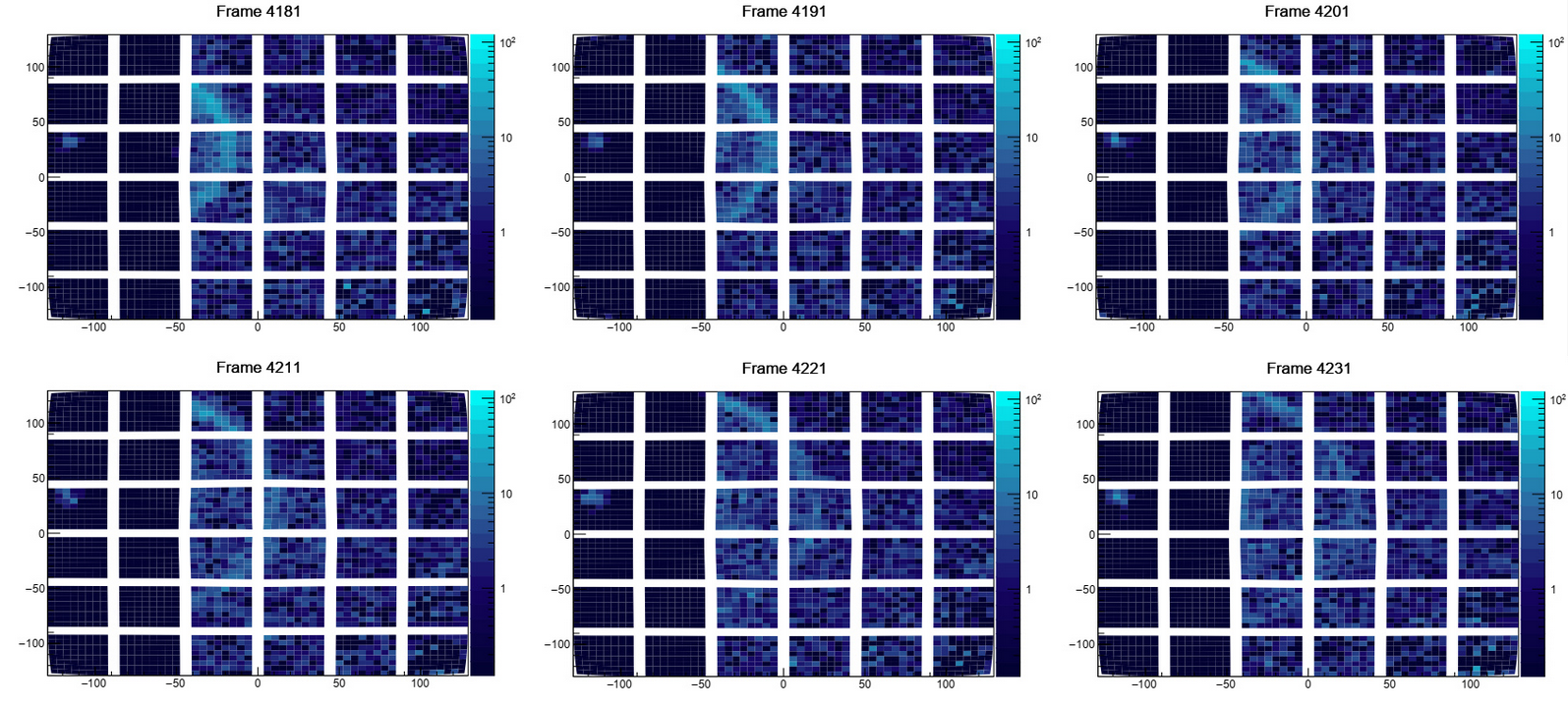}
\caption{A sample of  frames of an ELVE being observed in the focal surface of Mini-EUSO. Pictures are 10 frames apart,  corresponding to 5 $\mu$s (2.5 $\mu$s $\times$ 10). Note the dead area between different PMTs; some of the light between pixels is captured by the Winston cone (inverted pyramid) at the border of PMTs.}
\label{fig-ELVE}        
\end{figure}

In Figure \ref{fig-ELVE} are shown the pictures of a typical ELVE entering the field of view of Mini-EUSO as observed (on 2020/05/26) by the Focal Surface (FS). The ELVE is observed for $\simeq140$ GTUs, corresponding to $\simeq$320 $\mu$s. This event is peculiar because the lightning causing it (and thus the ELVE centre) is in the field of view (f.o.v) of the instrument (for all other events observed the centre of the event is outside the f.o.v.)  In this image sequence, the left column of three EC units (12 PMTs) is temporarily working at a reduced voltage (about 1/1000 sensitivity) due to a previous bright lightning activity that triggered the safety mechanism and caused the gain of these photomultipliers to be reduced. This allows to see the lightning in the centre left of the focal surface, which generates the ELVE ring expanding toward the centre and the right part of the FS.

{\bf Determination of ELVE centre}. To determine the coordinates of the centre of the ELVES, an iterative algorithm has been developed to fit ELVES’ rings with circles. Given a single time frame, the algorithm stores the positions of the detector’s pixels where an above-threshold signal was detected, recognizes if there is a circle arc to fit and gets the best fit parameters. The circle fitting problem is about detecting circles in images and computing their mathematical representation. As known, a circle of radius $R$ and centre coordinates $(a,b)$ is described by the equation $
(x-a)^2+(y-b)^2=R$. The goal of circle fitting algorithms is to calculate the parameters $(a,b,R)$ of the circle that best fits the image’s data points $(x_i,y_i)$, supposed to lie on a circle arc. Fitting curves is a notoriously difficult problem, and many efforts have been spent to develop a robust algorithm, which tolerates noisy images and gets the best fit parameters even for smaller (incomplete) arc data. Many algorithms have been developed for the fitting circle problem in the past. An algorithm, which is not sensitive to outliers and to data noise, and that allows to identify outliers and remove them, is the one proposed by Ladrón de Guevara \cite{Ladron}. The developed algorithm exploits the advantages of the circle fitting method proposed in \cite{Ladron}, and fits the data points by minimizing the sum of the geometric distances to the data points:  

\begin{equation}
F(a,b,R)=\sum_i \left[ \sqrt{\left( (x_i - a)^2 + (y_i-b)^2 \right) - R^2} \right]
\label{ErrFun}
\end{equation}

\begin{figure}[!htbp]
\centering
\includegraphics[width=0.35\textwidth]{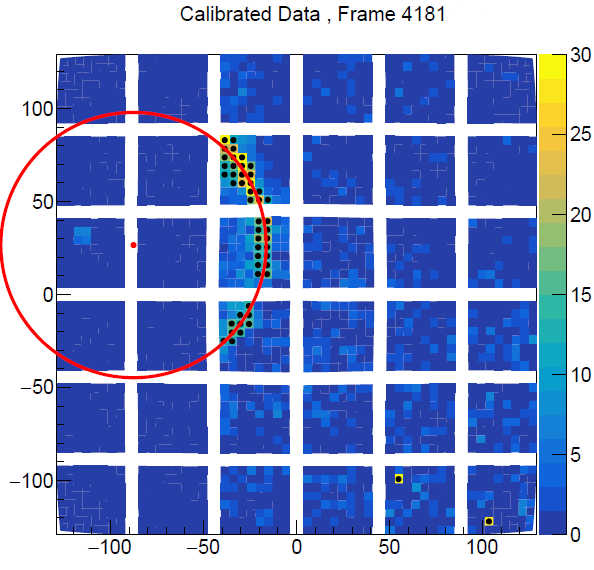}
\caption{Circle that best fits the data of one frame of an ELVE being observed by the focal surface of Mini-EUSO. The circle has its centre in $(-88,27)\; \text{km}$ and radius equal to $71 \;\text{km}$.}
\label{fig-FrameFitCircle}        
\end{figure}

The minimization of such function is performed by applying a steepest descent algorithm, which tries step by step to correct the prediction for the circle that best fits the starting data points, by moving the circle centre of an amount that depends on the position of the data points with respect to the predicted circle. The algorithm has been improved by introducing weight factors, so that each data point contributes to the displacement of the circle centre by an amount that depends not only on its position with respect to the circle, but also on the number of photo-electrons detected by that pixel. The optimal circle computed by the algorithm on one of the time frames of the ELVE event proposed above is depicted in Figure \ref{fig-FrameFitCircle}. The Mini-EUSO pixels whose coordinates were used to fit the circle are marked with black dots. The circle has its centre in $(-88,27) \;\text{km}$ and radius equal to $71 \;\text{km}$. Notice how the algorithm is able to find the circle that best fits the points even if there are some outliers in the bottom right corner of the image. Moreover, the circle passes through the pixels where a greater amount of photo-electrons was detected, as those pixels contribute most to the iterative computation of the best fit parameters. 

\begin{figure}[!htbp]
\centering
\includegraphics[width=0.55\textwidth]{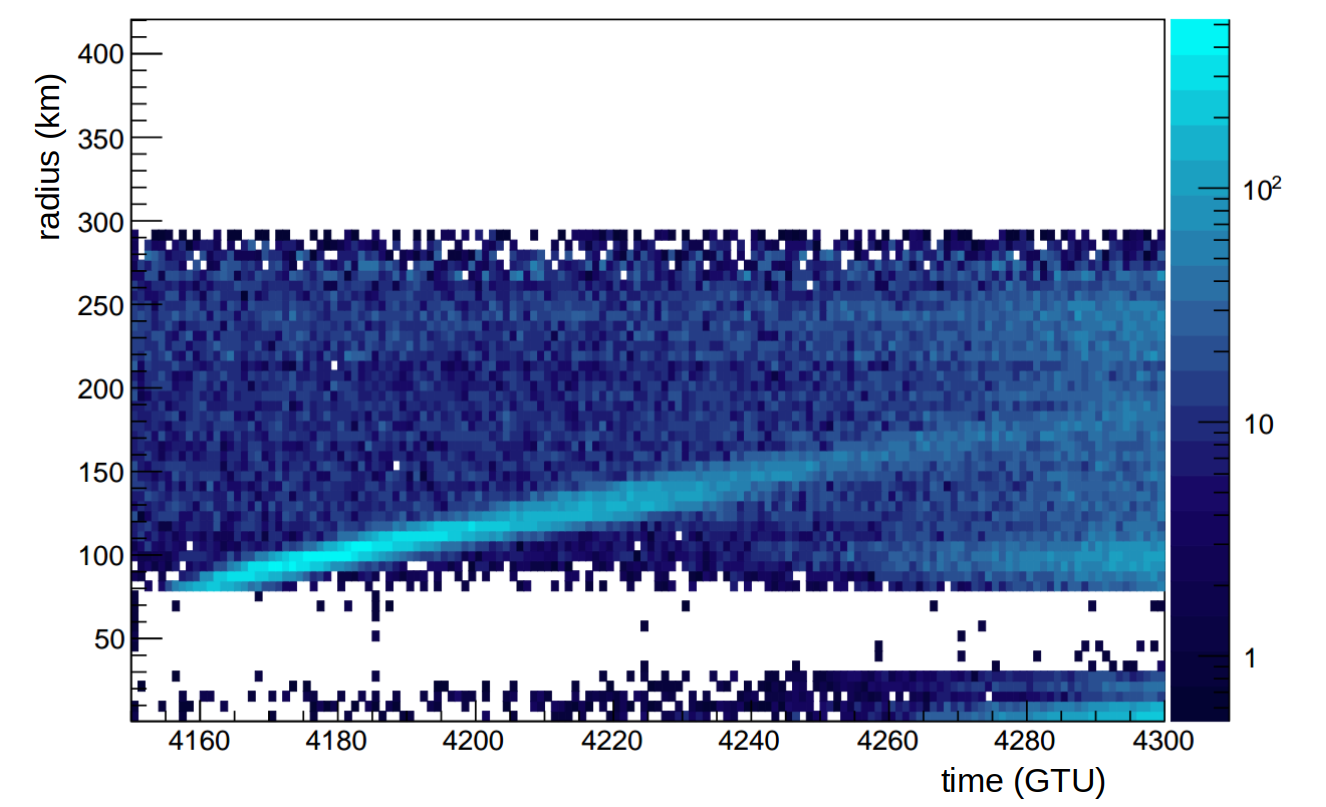}
\caption{Counts (Z-axis) as a function of radial distance R from the ELVE centre (Y-axis) and time (GTU, X-axis) for the ELVE measured with Mini-EUSO.}
\label{fig-ELVE-radius_vs_time}        
\end{figure}

\begin{figure}[!htbp]
\centering
\includegraphics[width=0.50\textwidth]{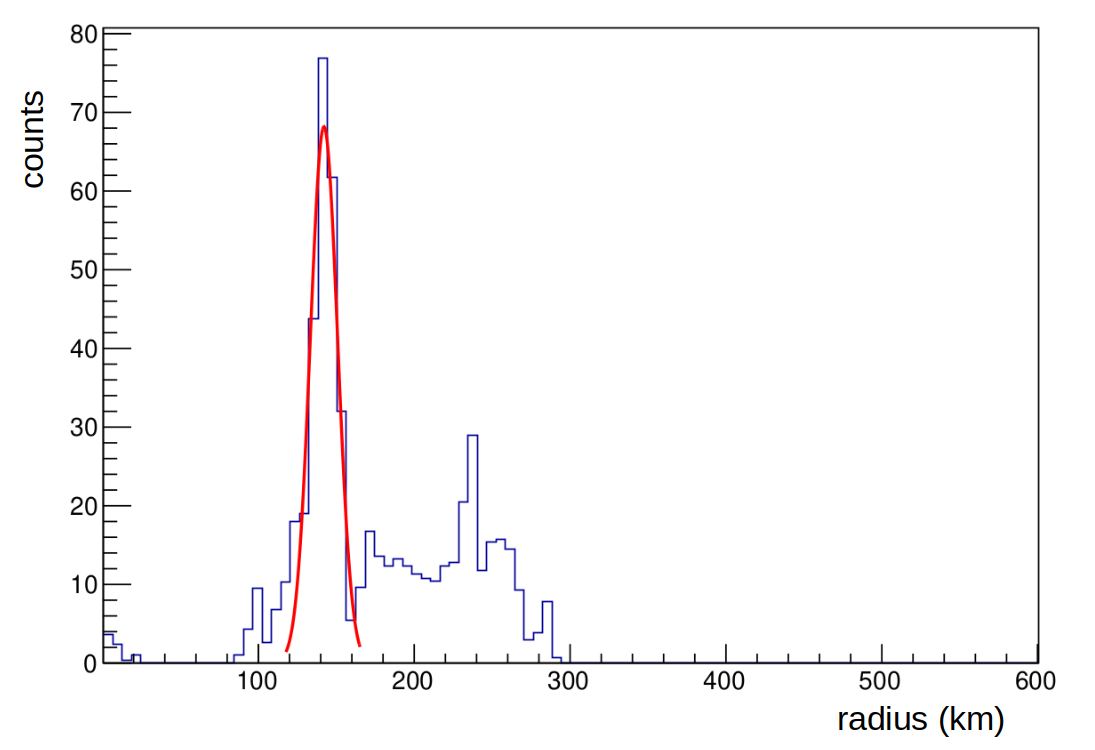}
\caption{Fitting of the total energy detected in the PDM in a single GTU (frame 4263) as function of the distance from the centre. X-axis: radial distance from the ELVE centre in km; Y-axis: number of counts at a given distance. The Gaussian fit is used to determine the radius-time relationship for subsequent analysis.}
\label{fig-ELVE-fit}        
\end{figure}

\begin{figure}[!htbp]
\centering
\includegraphics[width=0.50\textwidth]{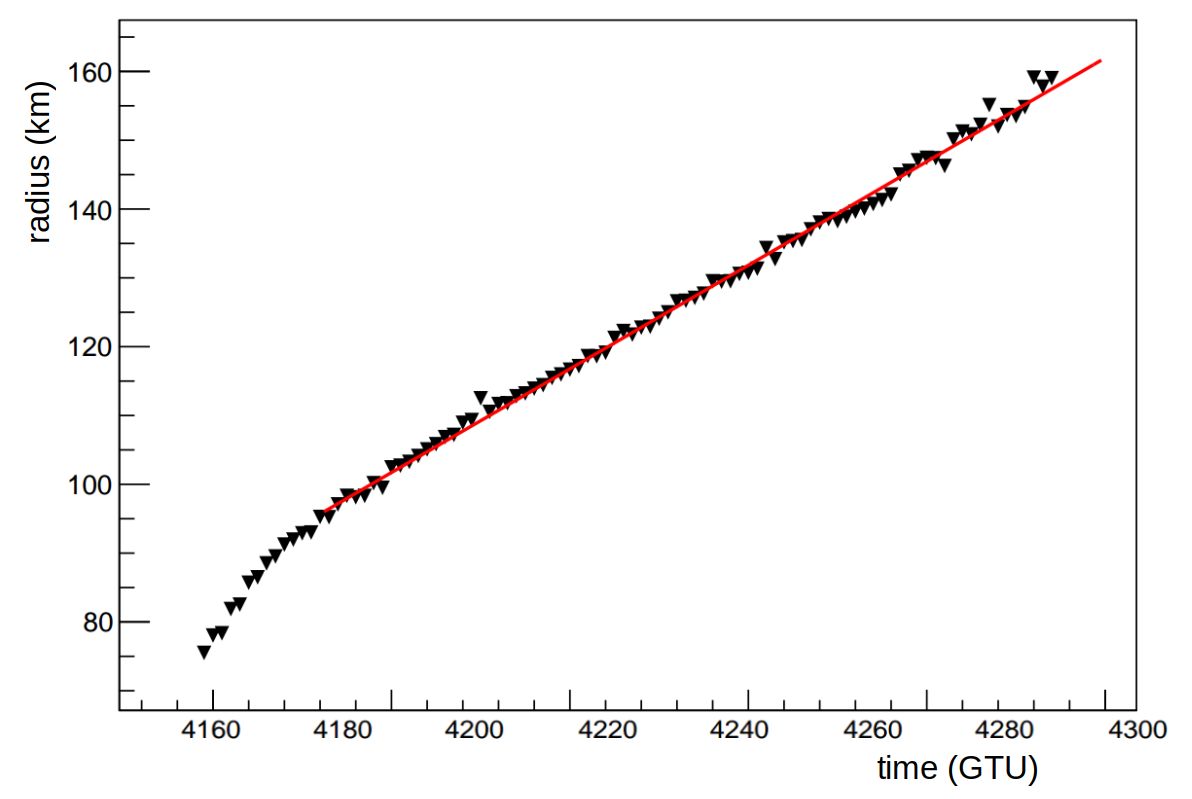}
\caption{Radius of the ELVE as a function of time. The linear fit gives the asymptotic speed of light; for the ELVE considered case we get: $v=301,000\pm 2,000$ km/s, consistent with $c$}
\label{fig-ELVE-radius}        
\end{figure}

{\bf Determination of arc parameters, radius and energy}. In Figure \ref{fig-ELVE-radius_vs_time} is shown the plot of the number of counts detected as a function of the radial distance  R (on the Y axis) at time t (the X axis). The ELVE can be seen as the high-count line, growing  as function of time (the horizontal band with R<80 km has fewer counts due to the aforementioned lower gain).
Fitting the  count  profile at each frame time with a Gaussian function we can determine the position of the radius of the ring and its thickness. As an example, we show in Figure \ref{fig-ELVE-fit} the projection on Y axis for the frame 4236 and the corresponding Gaussian fit. Note the very high signal/noise ratio for this event that allows to clearly identify the peak of the circle. The Radius-time expansion profile of the event, $R(t)$, is shown in Figure \ref{fig-ELVE-radius}. The simple geometry of a spherical expanding electromagnetic wave interacting with the plane geometry of the ionosphere result in the following equation:
\begin{equation}
R(t)=\sqrt{c^2t^2-(h-h_0)^2}
\end{equation}
where R(t) is the radius of the ELVE as function of time; $c$ is the speed of light; $h$ is the altitude of the ionosphere, $h_0$ is the altitude of the lightning /intra-cloud discharge that causes the event.

The velocity of apparent propagation of the ELVE is consequently given by:
\begin{equation}
V(t)=\frac{dR}{dt}=\frac{c^2t}{\sqrt{c^2t^2-(h-h_0)^2}}=\frac{c^2t}{R(t)}
\end{equation}

tending asymptotically to $c$ when $ct>>(h-h_0)$. 

In Figure \ref{fig-ELVE-radius} is also shown the linear fit of the asymptotic velocity of propagation; for the ELVE considered case we get: $v=301,000\pm 2,000$ km/s, consistent with $c$.

\section{Conclusions}\label{Conclusions}
In this work we have presented the observation capabilities of ELVES by Mini-EUSO detector.  Initial analysis of the  data received in the first year of operations confirms the correct functioning of the instrument and the high precision measurements of this class of TLEs can contribute to shed light on the various phenomena involved. Furthermore, in the near future it will be possible to make joint observations with other detectors on board the ISS such as Altea-Lidal \cite{lidal} and ASIM \cite{asim}.

\acknowledgments


This work was partially supported by Basic Science Interdisciplinary Research Projects of 
RIKEN and JSPS KAKENHI Grant (22340063, 23340081, and 24244042), by  the Italian Ministry of Foreign Affairs	and International Cooperation, by the Italian Space Agency through the ASI INFN agreements n. 2017-8-H.0 and n. 2021-8-HH.0, by NASA award 11-APRA-0058, 16-APROBES16-0023, 17-APRA17-0066, NNX17AJ82G, NNX13AH54G, 80NSSC18K0246, 80NSSC18K0473, 80NSSC19K0626, and 80NSSC18K0464 in the USA,   by the French space agency CNES, by the Deutsches Zentrum f\"ur Luft- und Raumfahrt, the Helmholtz Alliance for Astroparticle Physics funded by the Initiative and Networking Fund of the Helmholtz Association (Germany), by Slovak Academy of Sciences MVTS JEM-EUSO, by National Science Centre in Poland grants 2017/27/B/ST9/02162 and 2020/37/B/ST9/01821, by Deutsche Forschungsgemeinschaft (DFG, German Research Foundation) under Germany's Excellence Strategy - EXC-2094-390783311, by Mexican funding agencies PAPIIT-UNAM, CONACyT and the Mexican Space Agency (AEM), as well as VEGA grant agency project 2/0132/17, and by by State Space Corporation ROSCOSMOS and the Interdisciplinary Scientific and Educational School of Moscow University "Fundamental and Applied Space Research".

\clearpage
\section*{Full Authors List: \Coll\ JEM-EUSO}
%
%
%

G.~Abdellaoui$^{ah}$, 
S.~Abe$^{fq}$, 
J.H.~Adams Jr.$^{pd}$, 
D.~Allard$^{cb}$, 
G.~Alonso$^{md}$, 
L.~Anchordoqui$^{pe}$,
A.~Anzalone$^{eh,ed}$, 
E.~Arnone$^{ek,el}$,
K.~Asano$^{fe}$,
R.~Attallah$^{ac}$
% ...

\noindent
% ...
$^{ac}$ LPR at Dept. of Physics, Faculty of Sciences, University Badji Mokhtar, Annaba, Algeria \\
% ...
$^{ah}$ Telecom Lab., Faculty of Technology, University Abou Bekr Belkaid, Tlemcen, Algeria \\
% ...
$^{cb}$ Universit\'e de Paris, CNRS, AstroParticule et Cosmologie, F-75013 Paris, France\\
% ...
$^{ed}$ Istituto Nazionale di Fisica Nucleare - Sezione di Catania, Italy\\
% ...
$^{eh}$ INAF - Istituto di Astrofisica Spaziale e Fisica Cosmica di Palermo, Italy\\
% ...
$^{ek}$ Istituto Nazionale di Fisica Nucleare - Sezione di Torino, Italy\\
$^{el}$ Dipartimento di Fisica, Universita' di Torino, Italy\\
% ...
$^{fe}$ Institute for Cosmic Ray Research, University of Tokyo, Kashiwa, Japan\\ 
% ...
$^{fq}$ Nihon University Chiyoda, Tokyo, Japan\\ 
% ...
$^{md}$ Universidad Polit\'ecnia de madrid (UPM), Madrid, Spain\\
% ...
$^{pd}$ University of Alabama in Huntsville, Huntsville, AL; USA\\
$^{pe}$ Lehman College, City University of New York (CUNY), NY, USA\\
% ...


\begin{thebibliography}{99}
\bibitem{Mini-EUSO-Astrophys} S. Bacholle et al., ApJ Supp.,  253, 2, 36, 17, 2020.
\bibitem{capel}  F. Capel et al., J. Astron. Telesc. Instrum. Syst., 5, 044009, 2019.
\bibitem{matteo}  M. Battisti et al. (JEM-EUSO Coll.), Overview of the Mini-EUSO $\mu$ s trigger logic performance, PoS(ICRC2021) 411. 
\bibitem{giorgio}  G. Cambiè et al. (JEM-EUSO Coll.), Integration and qualification of the Mini-EUSO telescope on board the ISS, PoS(ICRC2021) 1001.
\bibitem{casolino} M. Casolino et al. (JEM-EUSO Coll.), The Mini-EUSO telescope on board the International Space Station: Launch and first results,  PoS(ICRC2021) 886.
\bibitem{lech}  L. Piotrowski et al. (JEM-EUSO Coll.), Towards observations of nuclearites with Mini-EUSO, PoS(ICRC2021) 1181.
\bibitem{kenji}  K. Shinozaki et al. (JEM-EUSO Coll.), Measurement of UV light emission of the nighttime Earth by Mini-EUSO for space-based UHECR observations, PoS(ICRC2021) 1165.
\bibitem{alessio} A. Golzio et al. (JEM-EUSO Coll.), A study on UV emission from clouds with Mini-EUSO, PoS(ICRC2021) 417.
\bibitem{francesco}  F. Fenu et al. (JEM-EUSO Coll.), Simulation studies for the Mini-EUSO detector, PoS(ICRC2021) 757.
\bibitem{inan91} U. S. Inan, T. F. Bell, J. V. Rodriguez, Geophysical Research Letters,  18,4, 705,1991.
\bibitem{doi:10.1029/91GL03168} W. L. Boeck, et al., Geophysical Research Letters, 19, 2, 99, 1992.
\bibitem{Fukunishi1996} H. Fukunishi et al., Geophysical Research Letters, 23, 2157, 1996.
\bibitem{Ladron} Ladrón de Guevara et al., J. Math. Imaging Vis., 40, 147–161, 2011.
\bibitem{lidal} A. Rizzo et al., Journal of Physics: Conference Series, 1226, 012024, 2019.
\bibitem{asim} N. Østgaard et al., Space Science Reviews, 215, 23, 2019.




\end{thebibliography}
\end{document}